\documentclass{mem_coronae}
\usepackage{natbib}\usepackage{txfonts}\usepackage{balance}
\usepackage{graphicx}
\usepackage[a4paper]{hyperref}
\idline{78 n.2}{1}
\begin{document}

\title{The origin of X-ray emission from T Tauri stars}

   \subtitle{}

\author{
Thomas Preibisch
          }

  \offprints{Th. Preibisch}

\institute{
 Max Planck Institute for Radio Astronomy,
 Auf dem H\"ugel 69, D-53121 Bonn,
 Germany
\email{preib@mpifr-bonn.mpg.de}
}

\authorrunning{Thomas Preibisch}

\titlerunning{The origin of X-ray emission from T Tauri stars}

\abstract{
Several aspects concerning the origin of the very strong 
X-ray activity of 
T Tauri Stars (TTS) are still not well understood.
Important new insight came recently from the
$Chandra$ Orion Ultradeep Project (COUP), 
a unique 10-day long $Chandra$ observation of the Orion
Nebula Cluster,
and the XMM-Newton Extended Survey of the Taurus Molecular 
Cloud (XEST).
Based mainly on the results of these two large projects,
I will discuss our current knowledge about the
location of the X-ray emitting structures in TTS,
the nature of their coronal magnetic fields,
inferences for pre-main-sequence magnetic dynamos, and
the relation between accretion processes and X-ray emission.
}
\maketitle{}

\section{Introduction}

T Tauri stars (TTS) are low-mass ($M \leq 2\,M_\odot$) 
pre-main sequence stars with typical ages between $\la 1$~Myr
and a few Myr.
They come in two flavors: the classical T Tauri stars (CTTS)
show H$\alpha$ emission and infrared excesses, which are a signature
of circumstellar disks from which the stars  accrete
matter. The weak-line T Tauri stars (WTTS), on the other hand,
have already lost (most of) their circumstellar material 
and show no evidence of accretion.
TTS generally show highly elevated levels of X-ray 
activity, with
X-ray luminosities up to $\sim 10^4$ times and 
plasma temperatures up to $\sim 50$ times higher than seen in our 
Sun \citep[e.g.,][]{FeigelsonMontmerle99}.
This strong X-ray emission 
has far-reaching implications
for the physical processes in the circumstellar environment, the
formation of planetary systems, and the evolution of 
protoplanetary atmospheres
\citep[e.g.,][]{Glassgold05,Wolk05}.

After the first discoveries of X-ray emission from TTS
 with the EINSTEIN satellite
\citep[e.g.,][]{Feigelson81}, many star forming regions and young
clusters have been 
observed with different X-ray observatories
\citep[e.g.,][]{Casanova95,Gagne95,Preibisch96,Feigelson02,Preibisch02,Flaccomio03}.
While these observations provided important information
about the X-ray properties of TTS, there were also serious
limitations.
First, the typical samples of X-ray detected objects in each
observation 
contained hardly more than $\sim 100$ objects, too few to allow
well founded statistical conclusions to be drawn.
Second, a large fraction of the known cluster members (especially
low-mass stars) remained undetected
in X-rays, and any correlation studies had therefore to deal with
large numbers of upper limits.
Third, especially in dense clusters, the individual sources could
often not be
spatially resolved, and so the proper identification of the X-ray
sources was
difficult or impossible.
Finally, in most X-ray data sets, 
only a relatively small number of individual young stars
were bright enough in X-rays to allow their spectral and temporal
X-ray properties to be studied in detail, and it was not clear
whether these stars
really are ``typical'' cases or perhaps peculiar objects.

The basic, still unresolved question concerns the exact
origin of the X-ray activity of TTS. 
Although there is strong evidence that in most TTS
the X-ray emission is related to coronal magnetic activity,
it is unclear what kind of structures may be the
building blocks of TTS coronae and whether
these coronae are created and heated by
solar-like (although strongly enhanced) magnetic dynamo processes,
or whether different kinds of  magnetic structures
and heating mechanisms are involved.
Furthermore, a fundamentally different source of X-ray emission
may be present in actively accreting TTS:
the shocks where the accreted material
crashes onto the stellar surface seem to produce 
soft X-ray emission in some TTS \citep[see, e.g.,][]{Kastner02}.
Hot ($\ga 10 - 30$~MK)
coronal plasma may coexist with cool ($\la 1-3$~MK) plasma
in accretion shocks \citep{Schmitt05}.
An important question, therefore, is whether 
accretion shocks are an important source of TTS
X-ray emission or only relevant in a few, perhaps  peculiar, objects.

\section{Large X-ray projects on TTS}

Very significant progress on these and other questions
has been made in the last few years
with two major observational projects that provided unprecedented
X-ray data sets on TTS.

The first one is the $Chandra$ Orion Ultradeep Project (COUP), 
a unique, 10-day long (total exposure time of 838\,100 sec)
observation of the Orion Nebula Cluster (ONC)
with $Chandra$/ACIS \citep[for details of the observation and
data analysis see][]{Getman05}. 
This is the deepest and longest X-ray 
observation ever made of a young stellar cluster and 
produced the most comprehensive dataset ever acquired on the X-ray
emission of young stars. 
Nearly all of the 1616 detected X-ray sources could be 
unambiguously identified with optical or near-infrared counterparts.
With a detection limit of 
$L_{\rm X,min} \sim 10^{27.3}$~erg/sec for lightly absorbed 
sources, X-ray emission from more than 97\% of the 
$\sim 600$ 
optically visible and well characterized late-type 
(spectral types F to M) cluster stars
was detected \citep{Preibisch_coup_orig}; as the remaining $<3\%$ undetected stars are probably
no cluster members but unrelated field stars, the
COUP TTS sample is {\em complete}.

The other large project is the
XMM-Newton Extended Survey of the Taurus Molecular Cloud (XEST), 
a survey of the densest stellar populations of the 
Taurus Molecular Cloud, 
in X-rays and in the near ultraviolet \citep[for details, see][]{Guedel07}.
The principal data were extracted from 28 different XMM-Newton
exposures with the EPIC cameras,
covering a total of 5 square degrees,
and provided X-ray data on 110 optically well characterized TTS.
For several bright objects, high-resolution X-ray spectra
were obtained with the Reflection Grating Spectrometers.

The main papers discussing the origin of TTS X-ray emission 
are \citet{Preibisch_coup_orig} for COUP
and \citet{Briggs07} for XEST.
Note that
several of the results discussed below were already suspected 
from the data of shorter X-ray observations of different
star forming regions,
but were confirmed with better data quality and 
much higher statistical power 
in the COUP and XEST data sets.

\section{Some general results}
\begin{figure*}
\resizebox{\hsize}{!}{\includegraphics[clip=true]{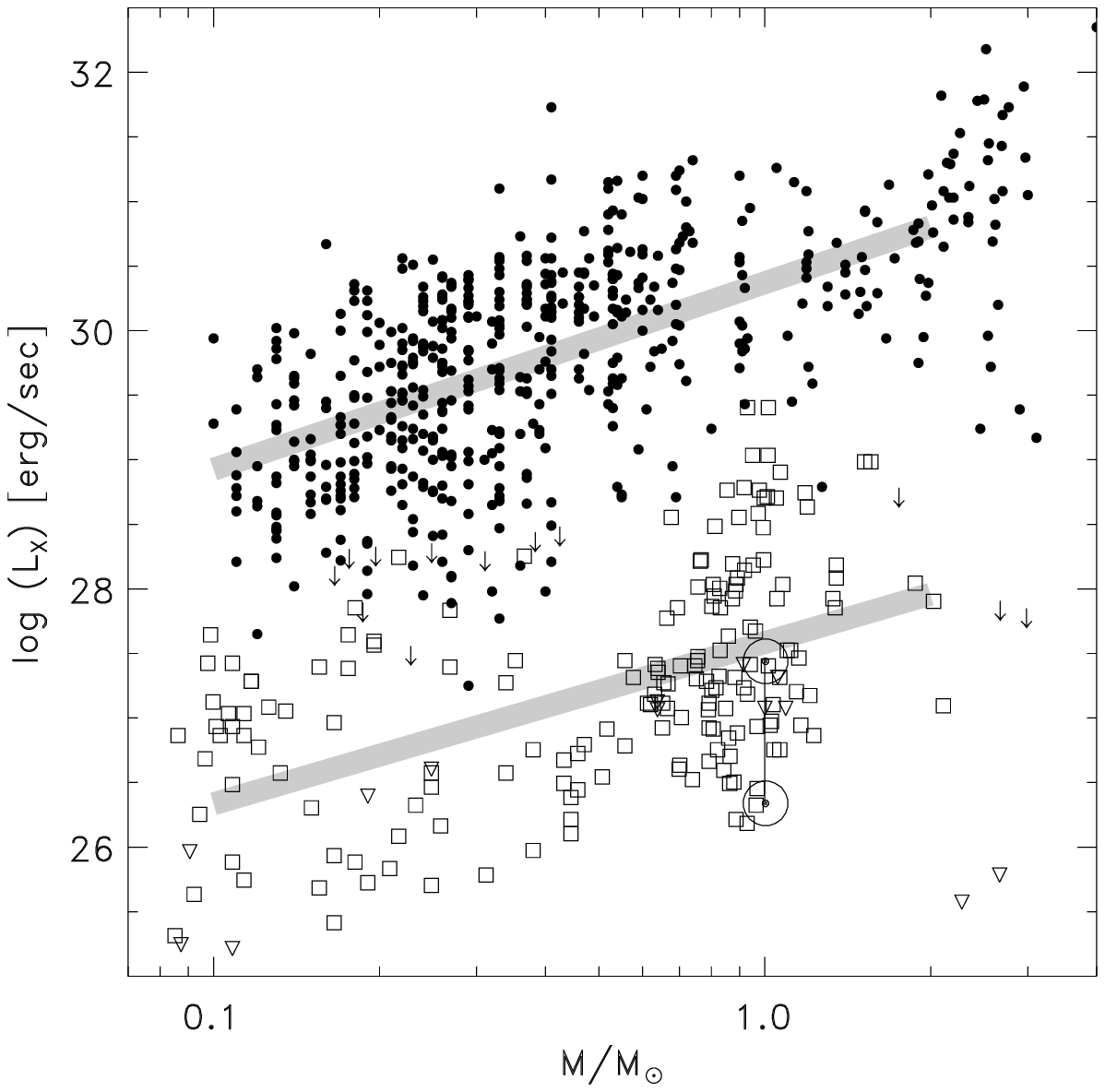}
\includegraphics[clip=true]{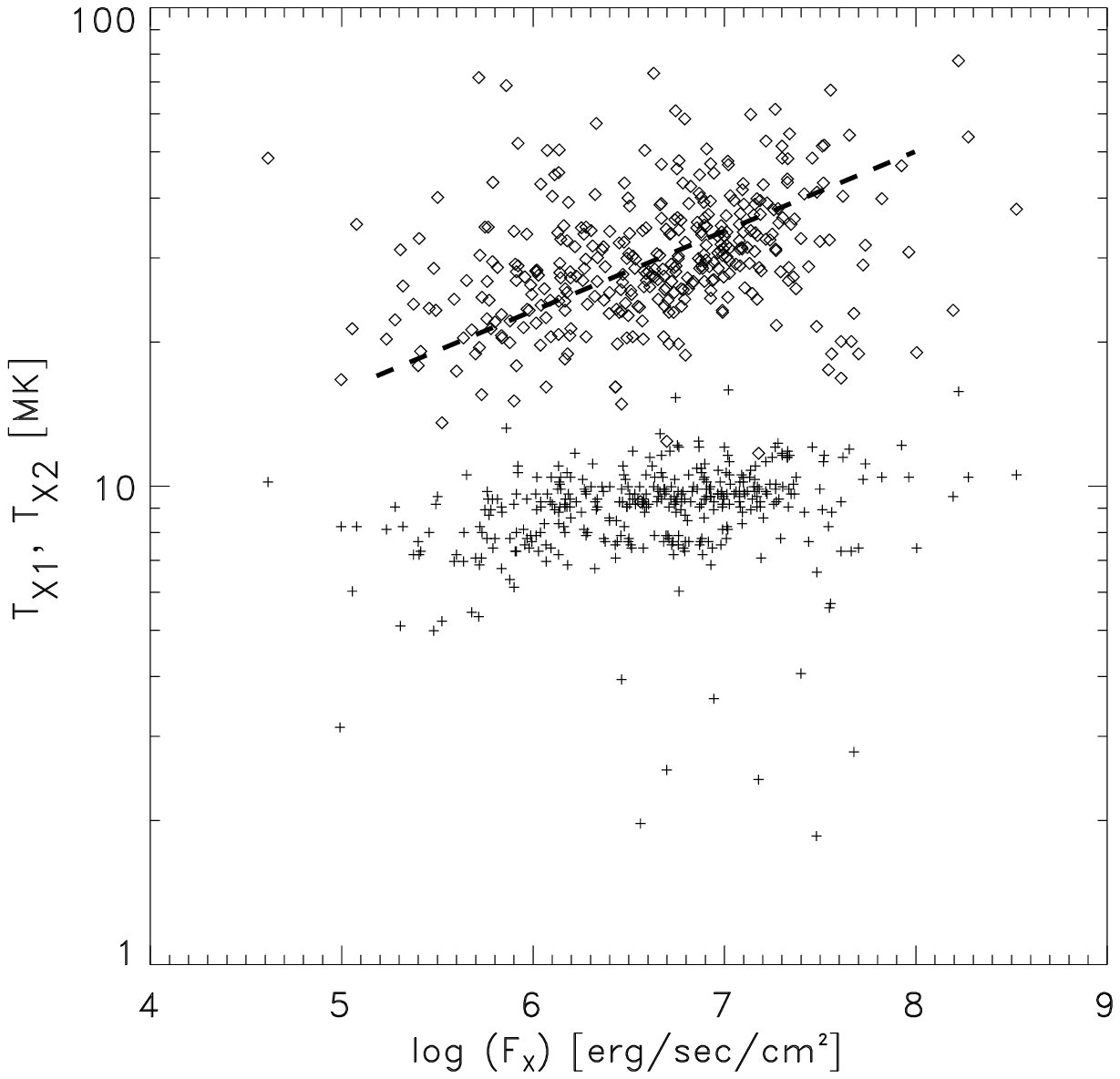}}
\caption{\footnotesize
{\bf Left:} X-ray luminosity versus stellar mass for the stars
in the COUP optical sample (solid dots, arrows for upper limits)
and for the NEXXUS sample of nearby field stars
(open squares, triangles for upper limits).
The thick grey lines show linear regression fits
for the low-mass ($M \leq 2\,M_\odot$) stars in these two samples.
{\bf Right:} Plasma temperatures (crosses for the cool component 
$T_{\rm X1}$, diamonds
for the hot component $T_{\rm X2}$) derived in the X-ray spectral fits
for the TTS in the COUP optical sample
plotted versus the X-ray surface flux.
The dashed line shows the
relation $F_X \propto T^6$.
\label{lx_m_tx_fx.fig}}
\end{figure*}

Nearly all TTS show
$L_{\rm X}/L_{\rm bol} > 10^{-5}$ and are therefore much more
X-ray active than the Sun 
($L_{\rm X,\odot}/L_{\rm bol,\odot} \sim 10^{-6}$). 
There is thus no indication for the existence of an ``X-ray quiet''
population of stars with suppressed magnetic activity.
The detection of X-ray emission from several 
spectroscopically-identified brown dwarfs
\citep[e.g.,][]{Preibisch_coup_bd} clearly shows that 
the X-ray activity does not terminate at the stellar mass limit
but extends well into the sub-stellar regime.

The X-ray luminosities of the TTS are correlated to stellar mass
(Fig.~\ref{lx_m_tx_fx.fig}, left)
with a power-law slope similar to that found for the NEXXUS
stars \citep{Schmitt04}, a complete sample of nearby 
late-type field stars.
The plasma temperatures of the COUP TTS 
derived in fits to the X-ray spectra with two-temperature models
are shown in the right panel of Fig.~\ref{lx_m_tx_fx.fig}.
The temperature of the hot plasma component
increases with increasing surface flux. 
The temperatures of the cool plasma
component of most TTS are remarkable similar and around 10~MK.

The TTS generally show high-amplitude rapid variability,
with typically one or two very powerful flares 
($L_{\rm X, peak} \ga 10^{30 \dots 32}\,\rm erg/sec$) per week on 
each star.


\section{X-ray emission and accretion}

\subsection{Is X-ray emission from accretion shocks important ?}

According to the magnetospheric accretion
scenario, accreted material crashes onto the stellar surface with
velocities of up to several 100 km/sec, what should cause
shocks with temperatures of up to about $\sim 10^6$~K,
in which strong optical and UV excess emission
and perhaps also soft X-ray emission is produced.
The expected characteristics of
X-ray emission from accretion shocks would be a very soft spectrum
(due to the low plasma temperature in the shock), 
and perhaps simultaneous brightness variations at optical/UV
wavelengths and in the X-ray band. 
Recent high-resolution X-ray spectroscopy of {\em some}
TTS  \citep[e.g.~TW Hya, XZ Tau
and BP Tau, see][]{Kastner02, Favata03, Schmitt05}
yielded very high electron densities   
($n_e \sim 10^{13}\,\rm cm^{-3}$) in the
coolest (1\dots 5~MK), O\,VII and Ne\,IX  forming plasma components,
what
has been interpreted as evidence for X-ray emission originating from
accretion shocks (rather than coronal loops, with their
typical densities
of $n_e \sim 10^{9\dots 11}\,\rm cm^{-3}$).

However,
neither the COUP nor the XEST results provided support for
a scenario in which the X-ray emission from TTS 
is dominated by accretion shocks. First,
the X-ray luminosities of many accreting TTS are {\em  larger}
than, or similar to, their accretion luminosities, ruling out
the possibility 
that the bulk of the observed X-ray emission from
the TTS could originate from accretion processes.
 
Second, the X-ray  spectra of nearly all TTS show much higher plasma
temperatures (typically a $\sim 10$~MK cool component and $\gtrsim
20$~MK hot component) than the $\lesssim 1-3$~MK expected from shocks
for the typical accretion infall velocities.
The vast majority of the TTS show 
neither significant plasma components at temperatures below
3~MK, nor indications for soft ($\la 1$~keV) excesses that may hint
towards emission from accretion shocks.

Third, the
high-resolution spectra of TTS analyzed in the XEST project
did not show  any evidence for the high plasma densities as
expected for accretion shocks;
the derived densities are only
$n_e \sim 3 \times 10^{11}\,\rm cm^{-3}$ for BP Tau 
and
$n_e < 10^{11}\,\rm cm^{-3}$ for 
T Tau N  and the Herbig star AB Aur
\citep{Telleschi07}. Such low densities are not compatible
with standard assumptions of accretion shocks.
 
Fourth, from simultaneous  X-ray and optical 
monitoring of 800 stars in the ONC,  \cite{Stassun06} 
found that 95\% of the ONC TTS did {\em not} show 
any time-correlated X-ray - optical modulations
that would be expected if surface accretion shocks
were the dominant sites of X-ray production.

These results show clearly that in the vast
majority of TTS the X-ray emission must be dominated by a
coronal component, and not by accretion shocks.
Of course, these arguments do not exclude the possibility
that accretion shocks may contribute {\em some fraction}
 of the X-ray emission in TTS.
It is critical to note that the CCD detectors  of $Chandra$  and 
XMM-Newton 
are not very sensitive to the cooler plasma expected from these
accretion shocks.
However, note that the
scenario of X-ray emitting accretion shocks also
faces problems from theoretical considerations.
The existence of accretion shocks does {\em not} necessarily 
imply that
one should expect {\em detectable} X-ray emission from these shocks:
according to models of the shock structure
\citep[e.g.,][]{Calvet98}, the material above the shock has typical
column densities of $\ga 10^{23}\,{\rm cm}^{-2}$ and
should thus completely absorb and thermalize the soft ($\la 0.5$~keV)
X-rays
emitted from the shock plasma within or close to the shock zone.
This problem has also been highlighted by \citet{Drake05}, who
argued that for the typically estimated  accretion rates  in
TTS ($\dot{M} \approx 10^{-7}\, M_\odot/{\rm yr}$),
the shock is buried too deeply in the
stellar atmosphere to allow the soft X-ray emission to escape
and be detected;
only for very low accretion rates ($\dot{M} \la 10^{-9}\,
M_\odot/{\rm yr}$) detectable soft X-ray emission can be expected.

\subsection{The suppression of X-ray emission by accretion}

The COUP data confirmed previous indications for systematic
differences in the X-ray properties of accreting and non-accreting TTS.
The absolute as well as the fractional X-ray luminosities
of  accreting TTS are systematically {\em lower} by a factor of
 $\sim 2-3$ than the corresponding values for  non-accreting TTS.
Also, X-ray activity appears to be anti-correlated
with mass accretion rate. These results were very well confirmed
with the XEST data and one can thus conclude that
the X-ray activity of accreting TTS is somehow suppressed.

The most likely explanation for this effect 
are changes in the coronal magnetic field 
structure by the accretion process. 
The pressure of the accreting material may distort the large-scale
stellar magnetic field \citep[e.g.][]{Romanova04} and
the magnetospheric transfer of material to the star can give rise to
instabilities of the magnetic fields around the inner disk edge.
The presence of accreting material should also lead to higher 
densities in (parts of) the magnetosphere; these high densities
may inhibit magnetic heating of the
accreting material to X-ray emitting temperatures.
The accreting material will also cool the corona when it
penetrates into active regions and mixes with hot plasma.
If the plasma gets cooled below a few MK,
its very soft X-ray emission is essentially undetectable for
the CCD X-ray detectors of $Chandra$ and XMM-Newton, and thus the
observed X-ray luminosity of the accreting stars is lower
than that of non-accretors 
\citep[see also][]{Telleschi07}.

\citet{Jardine06} have recently
modeled the X-ray emission of TTS assuming that
they have isothermal, magnetically confined coronae. 
In stars without a circumstellar disk,
these coronae extend outwards until the pressure of the
hot coronal gas overcomes the magnetic field, 
explaining the observed increase in the X-ray emission measure 
with increasing stellar mass.
In stars that are surrounded by a circumstellar accretion disk,
the outer parts of the coronal magnetic field are stripped
by the interaction with the disk.
This stripping provides a good explanation for 
the observed lower X-ray luminosities of accreting stars.


\section{X-ray emission from magnetic star-disk interactions?}

Another possibility for a non-solar like origin of the X-ray 
emission from TTS
may be plasma trapped in magnetic fields that connect the star with
its surrounding accretion disk.  The dipolar stellar magnetic field
lines anchored to the inner part of the accretion disk should be
twisted around because of the differential rotation between the star
of the disk. This twisting should lead to reconnection events
that heat the trapped plasma to very hot, X-ray emitting temperatures
and produce large X-ray flares.

\citet{Favata05} performed a detailed MHD model analysis  for the
$\sim 30$ largest flares 
seen in COUP data. The analysis suggests
that very long magnetic structures (more than a few times 
the stellar radius) actually
are present in {\em some} of the most active TTS.
Such very large structures may
indicate a magnetic link between these stars and their disks.
However, for the majority
of the analyzed flares much smaller loop lengths were found.
Furthermore, the COUP and XEST results
show that, in general, the X-ray luminosity is strongly linked
to stellar parameters like bolometric luminosity and mass, but does
not strongly depend on the presence or absence of circumstellar disks
as traced by near-infrared excess emission.
The bulk of the
observed X-ray emission from TTS therefore originates probably 
from more compact coronal structures, presumably 
 with geometries resembling
solar coronal fields.


\section{X-ray emission, rotation, and dynamos\label{rotation.sec}}

\subsection{X-ray activity and rotation}

\begin{figure*}
\resizebox{\hsize}{!}{\includegraphics[clip=true]{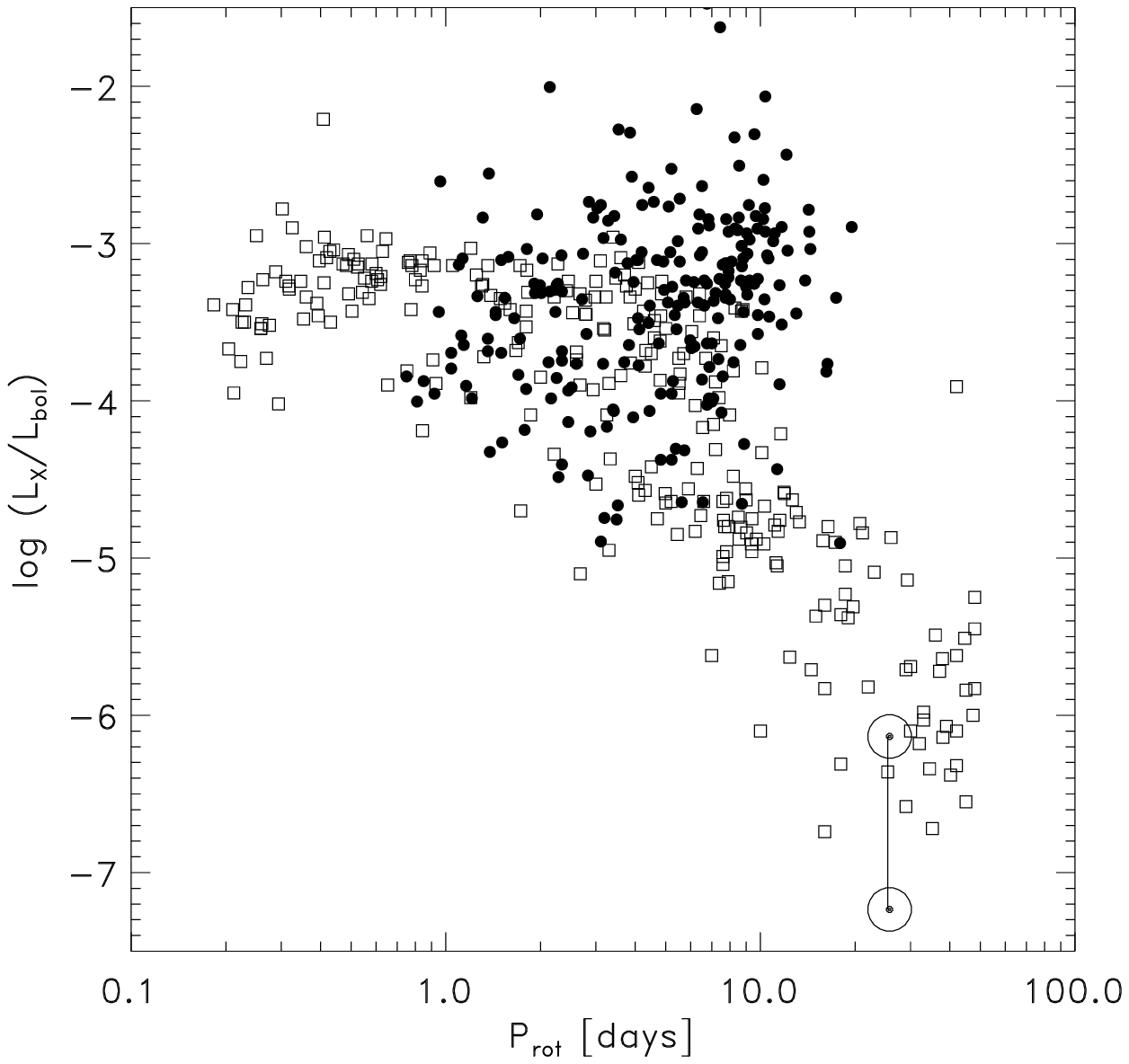}
\includegraphics[clip=true]{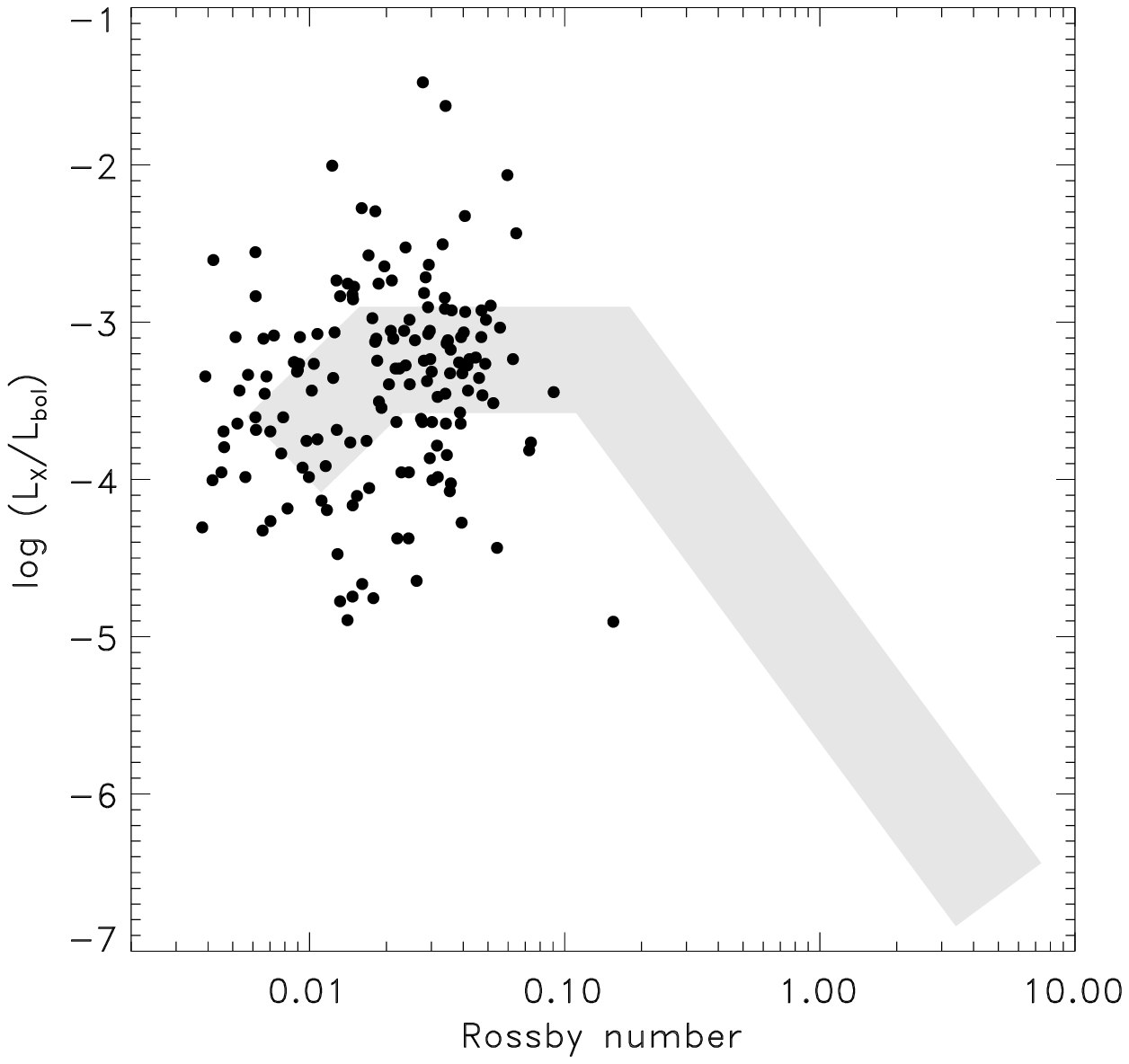}}
\caption{\footnotesize
{\bf Left:} Fractional X-ray luminosity versus rotation period.
This plot compares the ONC TTS (solid dots)
to data for main-sequence stars from \cite{Pizzolato03} and 
\cite{Messina03}
(open boxes) and the Sun.
{\bf Right:}
Fractional X-ray luminosity versus Rossby number
for the ONC TTS.
The grey shaded area shows the relation and the width of its
 typical scatter found for main-sequence stars.
\label{x_rot.fig}}
\end{figure*}

For main-sequence stars,
the well established correlation between
fractional X-ray luminosity and rotation period
\citep[e.g.][]{Pallavicini81, Pizzolato03} constitutes the main argument for
a solar-like dynamo mechanism as the origin of their X-ray activity.
The existence of a similar
relation between rotation and X-ray activity could never be
convincingly established for TTS; in most studies the small number of
X-ray detected TTS with known rotation periods did not allow to draw
sound conclusions.
A relation between rotation and X-ray activity was previously
suggested for the TTS in the Taurus star forming region 
\citep{Stelzer01}; however, new data have now revealed 
that this is only apparent because
the Taurus TTS population is biased toward fast rotators having, 
on average, higher
mass, thus being brighter in X-rays \citep{Briggs07}.
The COUP and the XEST data have both clearly confirmed 
that the TTS do {\em not} follow
the activity\,--\,rotation relation for main-sequence stars
(see Fig.~\ref{x_rot.fig}, left panel).

Theoretical studies of the solar-like $\alpha\!-\!\Omega$ dynamo
show that the dynamo number 
is not directly related to the rotation period, but to more complicated 
quantities such as the radial gradient of the angular velocity and the
characteristic scale length of convection  at the base of the
convection zone.
It can be shown that  the dynamo number 
is essentially proportional to the inverse square of the Rossby number $Ro$
\citep[e.g.][]{Maggio87}, which is 
defined as the ratio of the rotation period to the
convective turnover time $\tau_c$, i.e.~$Ro := P_{\rm rot}/\tau_c$.
For main-sequence stars, the theoretical expectations 
that the  stellar activity should show a tighter relationship to the
Rossby number than to rotation period
are well confirmed \citep[e.g.][]{Montesinos01}.
For large Rossby numbers, activity rises strongly as 
$L_{\rm X}/L_{\rm bol} \propto Ro^{-2}$ 
until saturation at $L_{\rm X}/L_{\rm bol} \sim 10^{-3}$ is reached
around $Ro \sim 0.1$,
which is followed by a regime of ``supersaturation'' 
for very small Rossby numbers, $Ro \lesssim 0.02$.

The convective turnover time scale is a sensitive function of 
the physical properties in the stellar interior.
The use of
semi-empirical interpolations of $\tau_c$ values as a function of,
e.g.,~$B-V$ color, may be appropriate for main-sequence stars,
but is clearly insufficient for TTS which have a very different and
quickly evolving internal structure.

In the analysis of the COUP data by \citet{Preibisch_coup_orig},
convective turnover times for the ONC TTS were  
computed from detailed stellar evolution models with the
Yale Stellar Evolution Code.
The right panel in Fig.~\ref{x_rot.fig} shows the 
fractional X-ray luminosities
of the ONC TTS versus the resulting Rossby numbers.
The plot shows no strong relation between these two quantities.
All TTS have Rossby numbers $< 0.2$ and therefore 
are in the saturated or super-saturated
regime of the activity -- Rossby number relation for main-sequence
 stars.
However,
a remarkable difference between the TTS and the 
main-sequence stars is apparent in the very wide dispersion of fractional 
X-ray luminosities
at a given Rossby number among the TTS. The scatter extends over
about three orders of magnitude and is 
in strong contrast to the tight relation found
for main-sequence stars, where the scatter in $\log\left(L_{\rm X}/L_{\rm bol}\right)$
at a given Rossby number is only about $\pm0.5$~dex \citep[e.g.,][]{Pizzolato03}.
This seems to suggest that additional factors, other than rotation,
 are important for the level of 
 X-ray activity in TTS.

\subsection{Implications for magnetic dynamos}

The activity-rotation relation shown by main-sequence
stars is usually interpreted 
in terms of the
$\alpha\!-\!\Omega$-type dynamo that is thought to work in the Sun.
Solar dynamo models assume that the
strong differential rotation in the tachocline, a region near the 
bottom of the convection zone in which the rotation
rate changes from being almost uniform in the radiative interior
to being latitude dependent in the convection zone,
generates strong toroidal magnetic fields.
While most of the toroidal magnetic flux is stored and further 
amplified in the tachocline, 
instabilities expel individual flux tubes, which then 
rise through the convection zone, driven by magnetic buoyancy, until 
they emerge at the surface as active regions.
The power of the dynamo (i.e.~the magnetic energy created 
by the dynamo per unit time) is
principally dependent on the
radial gradient of the angular velocity in the
tachocline and the characteristic scale length of convection at 
the base of the convection zone.
Faster rotating stars have stronger velocity shear in the
thin tachoclinal layer, causing the empirical relationship 
between X-ray luminosity and rotation rate in main-sequence stars.

Most TTS, however, are thought to be
fully convective, or nearly fully convective,
so the tachoclinal layer is either buried very deeply, or does not
exist at all.
Another kind of dynamo is thus required to explain the magnetic
activity of TTS.
Theoreticians have developed alternative dynamo concepts
\citep[e.g.][]{Durney93,Giampapa96,Kueker99,Dobler06}
that may work in fully convective stars. 
A general problem with these and other
models is that they disagree on
the type of large-scale magnetic topologies that
fully convective stars can generate, and that they
usually do not make quantitative predictions
that can be easily tested from observations.

Therefore, we once again consider the example of the Sun.
Although the solar coronal activity is most likely dominated 
by the tachoclinal dynamo action,
this does not prevent 
other dynamo processes from {\em also} operating. 
It is assumed that 
small scale turbulent dynamo action 
 is taking place throughout the solar convection
zone and is thought to be responsible for the small-scale intra-network fields.
This means that two conceptually distinct magnetic dynamos 
are simultaneously operating in the contemporary Sun,
although the solar
coronal activity is most likely dominated by the tachoclinal 
dynamo action.
It is therefore reasonable to assume that in the
(nearly) fully convective TTS, a convective dynamo is the
main source of the magnetic activity.

\section{Implications for coronal structure}

The up to $10^4$ times higher fractional X-ray luminosities
of TTS clearly require that the structure of their coronae must 
be quite different from
that of the Sun, where the X-ray emission is dominated by
a moderate number of active regions with magnetic field
configurations typically limited to heights of well below
one stellar radius. The coronae of TTS must be
either much more extended (at least several $R_\ast$)
or consist of structures with considerably higher plasma densities
than those on the Sun.

Various observational constraints are now available:
\citet{Flaccomio05} and \citet{Stassun06} used the COUP data
to search for
time-correlated X-ray - optical modulations in the ONC TTS.
More than 90\% of the TTS did not show such
time-correlated variability, what suggest a spatially rather 
homogenous 
distribution of X-ray emitting regions on the surface of the TTS.
On the other hand, some TTS did show apparently periodic
X-ray modulations with the same period as their rotation period
\citep{Flaccomio05}. This detection of rotational modulation in 
some TTS implies that the dominant X-ray emitting regions of 
these stars must be rather compact, distributed unevenly
around the star, and do not extend significantly more than a
stellar radius above the surface.

As mentioned above, the detailed MHD modeling of large flares
by \cite{Favata05} suggested that most of these flares 
occurred in rather compact loops ($l \la R_\ast$)
 with geometries resembling solar coronal fields.

Another (tentative) clue can be derived from the 
remarkable similarity of the temperatures ($\sim 10$\,MK)
of the cool plasma component in the COUP TTS sample.
This 10~MK component seems
to be a general feature of coronally active stars 
\citep[e.g.,][]{Sanz03} and 
may be related to a class of very compact loops
with high plasma density, presumably similar to X-ray bright points
on the Sun.


\section{Conclusions}

The observed X-ray properties of TTS strongly suggest that
the bulk of their X-ray emission has its origin in
coronal magnetic activity.
The surface of the TTS is probably covered
by a large number of compact and very dense magnetic structures,
which confine the X-ray emitting plasma.
Magnetic interaction between these regions may be
the driving source of the frequent and powerful X-ray flares.

In {\em some} TTS, very extended magnetic structure with lengths
of $\,>\!10 \times R_\ast$, which presumably connect the star to 
the circumstellar disk, seem to be involved.

The TTS do not follow the activity-rotation relation seen in
late-type main-sequence stars and the action of a solar-like
$\alpha\!-\!\Omega$\,-type dynamo seems to be excluded by their
(nearly) fully convective stellar structure.
The ultimate origin
of the X-ray activity of the TTS may be
a turbulent dynamo working in the stellar convection zone.

Accretion shocks at
the stellar surface can not be responsible for the
bulk of the observed X-ray emission in the vast majority of TTS.
Despite observational hints towards accretion shock related
X-ray emission in some TTS, this emission mechanism seems
important (in comparison to coronal emission) only in a few
exceptional objects.

\begin{acknowledgements}
I would like to thank Hans Zinnecker for many years of
motivation and advice in studying the X-ray emission of T Tauri stars.

\end{acknowledgements}

\bibliographystyle{aa}

\end{document}